\documentclass[useAMS,usenatbib]{mn2e}
\usepackage{amsmath,amssymb,wasysym,graphicx,lscape,natbib,epsfig,setspace,longtable,colortbl}
\usepackage{multirow}
\usepackage{booktabs}
\usepackage{lscape}

\usepackage{relsize}

\usepackage{fixltx2e}

\usepackage{amssymb}

\newcommand{\hi}{H\,{\sc i}}

\newcommand{\prim}{$^{\prime}$}
\newcommand{\prin}{$^{\prime\prime}$}
\newcommand{\aprox}{${\sim}$}

\newcommand{\km}{~km~s$^{-1}$}
\newcommand{\degree}{$^{\circ}$}
\newcommand{\halpha}{H${\alpha}$}
\newcommand{\cg}{CGCG\,097-}
\newcommand{\msolar}{M$_{\odot}$}

\newcommand{\msolaryr}{M$_{\odot}$\,yr$^{-1}$}

\newcommand {\apgt} {\ {\raise-.5ex\hbox{$\buildrel>\over\sim$}}\ }
\newcommand {\aplt} {\ {\raise-.5ex\hbox{$\buildrel<\over\sim$}}\ }

\setlongtables

\title[]{ Two long \hi\ tails in the outskirts of Abell\,1367}
\author[]{T. C. Scott$^{1,2,}$\thanks{E-mail:
tom@iaa.es (TS)}, L. Cortese$^{3}$,  E. Brinks$^{2}$, H. Bravo--Alfaro$^{4}$, R. Auld$^{5}$, 
\and   and R.  Minchin$^{6}$
\\
$^{1}$Instituto de Astrof\'\i sica de Andaluc\'\i a (CSIC),C/ Camino Bajo de Hu$\acute{e}$tor, 50, 18008 Granada, Spain\\
$^{2}$Centre for Astrophysics Research, University of Hertfordshire, College
Lane, Hatfield, AL10 9AB, UK\\ 
$^{3}$European Southern Observatory, Karl-Schwarzschild Str. 2, 85748 Garching bei Muenchen, Germany\\
$^{4}$Departamento de Astronom\'\i a, Universidad de Guanajuato,
Apdo, Postal 144, Guanajuato 36000, Mexico\\
$^{5}$School of Physics and Astronomy, Cardiff University, Cardiff CF24 3AA,
UK \\ 
$^{6}$NAIC-Arecibo Observatory, HC3 Box 53995, Arecibo, PR 00612, USA\\
}

\begin{document}
\date{Accepted. Received ; in original form }


\maketitle

\label{firstpage}

\begin{abstract}
We present VLA D--array \hi\  observations of the RSCG42 and FGC1287 galaxy groups, in the outskirts of the
Abell 1367 cluster. These groups are projected $\sim$ 1.8 and 2.7 Mpc west from the cluster centre.
\textcolor{black}{ The Arecibo Galaxy Environment survey provided evidence for \hi\ extending over as much as 200\,kpc in both groups.} Our new, higher resolution observations reveal that the complex \hi\  features detected by Arecibo are
in reality two extraordinary long \hi\ tails extending for $\sim$160 and 250 kpc, respectively, i.e.,
among the longest \hi\ structures ever observed in groups of galaxies.
Although in the case of RSCG42 the morphology and dynamics of the \hi\  tail, as well as the optical
properties of the group members, support a low-velocity tidal interaction scenario,
less clear is the origin of the unique features associated with FGC1287.
This galaxy displays an exceptionally long `dog leg' \hi\  tail and the large distance from the X--ray emitting
region of Abell\,1367 makes a ram-pressure stripping scenario highly unlikely. At the same
time a low-velocity tidal interaction seems unable to explain the extraordinary length
of the tail and the lack of any sign of disturbance in the optical properties of FGC1287.
An intriguing possibility could be that this galaxy might have recently experienced a high--speed interaction
with another member of the Coma--Abell\,1367 Great Wall. We searched for the interloper responsible
for this feature and, although we find a possible candidate, we show that without additional observations
it is impossible to settle this issue. While the mechanism responsible for
this extraordinary \hi\  tail remains to be determined, our discovery highlights how little we know about
environmental effects in galaxy groups.

\end{abstract}


\section{INTRODUCTION}

 \hi\ tails and streams represent some of the clearest observational evidence of the effect of the environment on the gas content of galaxies. It is now well established that several environmental mechanisms can be responsible for such remarkable features. Late-type, gas--rich systems entering the central region of clusters (within a radius of $\sim$1\, Mpc of the cluster centre) can have their gas stripped by ram pressure, P$_\mathrm{ram}$ = $\rho_\mathrm{ICM}$v$_\mathrm{rel}^2$ \citep{vgork04,roed07}. Several examples have been found in Virgo \citep{chung07} as well as in Abell\,1367 where \citet{scott10} report a 70 kpc tail emanating from \cg087. Moreover, \citet{ooster05} presented a 110\,kpc \hi\ structure emanating from NGC\,4388, proposing that it is a result of ram pressure stripping although not by the hot intracluster gas centred on M87, but rather by the hot halo gas of M86. 
 
 \hi\ tails can also be generated, under suitable initial conditions, from tidal interactions. A pair of stellar tails are expected to result from a tidal interaction and depending on the gas content of the progenitors, one or both of the tails may have a gaseous complement \citep{struck99}. In galaxy groups the lower relative velocities ($<$ 500 \km) favour slow tidal interactions that can displace a large fraction of the \hi\ from the parent galaxies and lead to lasting changes in their morphology. \textcolor{black}{Bekki et al. (2005a) \nocite{bekki05a} provide models in which they describe the formation of massive \hi\ clouds with no optical counterparts as high--density regions of intragroup \hi\ rings and arcs.} In Compact Groups such as Stephan's Quintet interactions can cause as much as 60\% of the group's \hi\ to be found in tidal tails and bridges  \citep{williams91}. Even in clusters of galaxies, high--speed interactions can produce incredibly long \hi\ tails. 
 \textcolor{black}{The best example is probably the case of NGC\,4254 \citep{minchin05,minchin07,hayn07}. Simulations   (Bekki et al. 2005b; Duc et al. 2008)  \nocite{bekki05b,duc08} suggests that the \hi\ stream in this system is  tidal debris from an interaction.}
 
As explained by \cite{hayn07}, the location of NGC\,4254 at $\sim$ 1Mpc from M87 argues against ram pressure being the main cause of this tail. Instead, they claim that a more likely cause is a tidal interaction, of the type expected under the harassment scenario proposed by \citet{moore96}. This idea was taken further by \citet{duc08}. The latter authors presenting a careful modelling of the tail originating from NGC\,4254 leading them to propose a high--speed interaction with the more massive galaxy NGC\,4192 (M98) to be at the source of the huge tail. 

All these observational studies highlight the importance of \hi\ tails for our understanding of environmental effects. Usually, the information provided by \hi\ observations, such as velocity, mass and \hi\ distribution provides crucial constraints for theoretical predictions of different environmental mechanisms \citep{voll03,duc08}. 

In order to unveil the main environmental mechanisms driving galaxy evolution in nearby clusters, we are conducting an \hi, molecular gas, and star formation study of late--type galaxies in the spiral-rich cluster Abell\,1367  \citep[][Scott et al. in preparation]{scott10}.  Abell\,1367 (z=0.022) lies at about the same distance as Coma, but with only about half the Coma ICM mass, and consists of two approximately equal--mass sub--clusters (SE and NW) that are in the process of merging \citep{cort04}. We are particularly interested to know if the young dynamical state of Abell\,1367 and its lower ICM content has any bearing on the relative roles of ram pressure and tidal interactions in driving the evolution of its spirals. The Arecibo Galaxy Environment Survey  \citep[AGES;][]{cort08}  survey revealed \textcolor{black}{\hi\ extending over up to 200\,kpc} in two groups, RSCG\,42 and FGC\,1287, located in projection beyond the cluster's 1.1 degree virial radius \citep[][]{rines03}.  This letter reports on follow-up VLA--D array \hi\ mapping of the galaxies in these groups. 

Based on a redshift to the cluster of 0.022 and assuming $\Omega_M =0.3$, $\Omega_\Lambda =0.7$, and $H_0$ =72\,\km\,Mpc$^{-1}$ \citep{sperg07} the distance to the cluster is 92\,Mpc with an  angular scale of 1\,arcmin $\approx$ 24.8\,kpc. The virial radius then corresponds to $\sim$ 1.64\,Mpc. We use J2000.0 coordinates throughout. Section \ref{secobs} details the observations, observational results are given in section  \ref{results} with discussion and concluding remarks in section \ref{4:conclusion}. 

\section{OBSERVATIONS}
\label{secobs}
\label{4obs}
\begin{figure*}
\begin{center}
\includegraphics[ angle=0,scale=0.7] {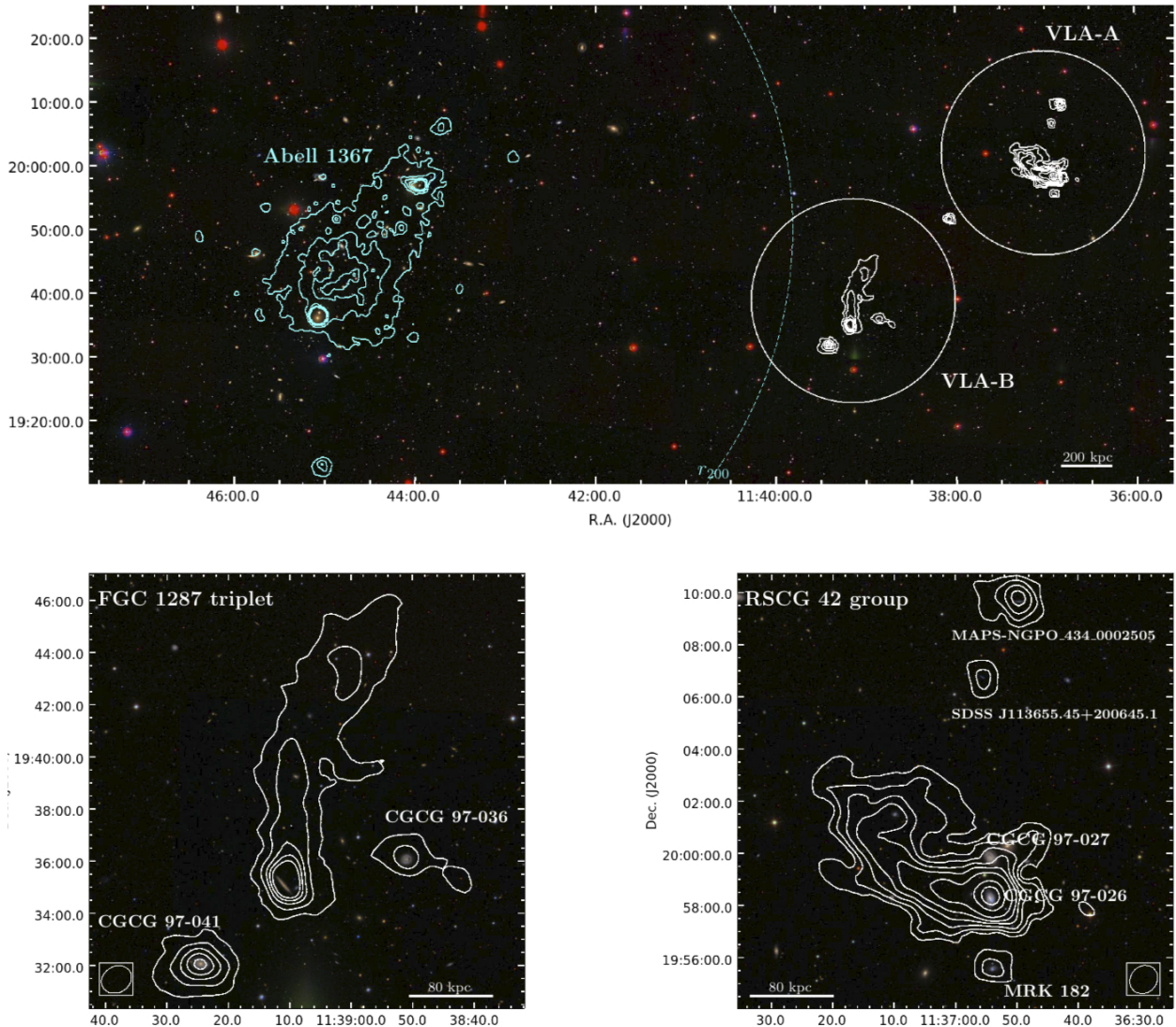}
\vspace{1cm}
\caption[]{\textbf{\textit{Upper panel:}}  Relative location of the RSCG\,42 (VLA --A) and FGC\,1287 (VLA--B) groups with respect to the Abell\,1367 cluster centre. The FWHM of the primary beam (32 arcmin) for each of the VLA D--array  fields  is indicated  with a white circle.   X--ray emission (\textit{ROSAT}) from the cluster ICM is indicated with cyan contours. The cluster's virial radius (1.64 Mpc $\sim$ 1.1\degree) is indicated with a dashed cyan chord.  \textbf{\textit{Lower left panel}}   FGC\,1287 triplet: white contours trace the natural weight  \hi\ surface density, with the outer contour indicating a column density of  \textit{N}$_{HI}$  = 2.0 x 10$^{19}$ cm$^{-2}$, with higher levels at (5, 10, 15, 20 and 40) x 10$^{19}$ cm$^{-2}$.  \textbf{\textit{Lower right panel}} RSCG\,42 compact group:  the contours trace the natural weight  \hi\ surface density, with the outer contour indicating a column density of  \textit{N}$_{HI}$  = 1.7 x 10$^{19}$ cm$^{-2}$, with higher levels at (3, 6, 10, 15, 25, 35, 50  and 70)  x 10$^{19}$ cm$^{-2}$. In all three panels the first \hi\ contour  corresponds to a 3 $\sigma$ detection in two channels.  The contours are overlaid on SDSS $u, g$ and $r$--band composite images. In  the lower panels the size of the D--array synthesised beam is indicated with the  white boxed ellipse. }
\label{fig1}
\end{center}
\end{figure*}

\hi\ was observed in two fields (VLA--A and VLA--B) with the NRAO\footnote{The National Radio Astronomy Observatory is a facility of the National Science Foundation operated under cooperative agreement by Associated Universities, Inc.} VLA in D--array  configuration with $\sim$2.5\,hr integrations.  The FWHM (32 arcmin) of the VLA primary beams for each of the observed  fields (large white circles) and  the intensity of the X--ray emission ({\em ROSAT}) from the cluster's ICM (cyan contours) are shown in the top panel of Figure \ref{fig1}.  In addition we made two follow-up pointings with the Arecebo 305--m telescope\footnote{The Arecibo Observatory is part of the National Astronomy and Ionosphere Center, which is operated by Cornell University under a cooperative agreement with the National Science Foundation.}. The observational set up for each of the VLA fields is listed in Table \ref{4:tt1}, including integration time, velocity resolution and  velocity range.

The observations were made on April 16, 2007. The data were calibrated and imaged with the AIPS software package, but the standard calibration and reduction procedures had to be adapted to overcome a number of issues associated with the VLA--to--EVLA transition. For both fields self--calibration (phase only) was carried out to mitigate the effects of side lobes from strong continuum sources. The continuum was fitted in line--free channels and subtracted  from the cubes using the AIPS task {\sc uvlsf}. Two of the observed galaxies have long, low surface brightness \hi\ tails.  In order to highlight the extent of these tails we applied natural weighting resulting in a synthesised beam of $\sim$65 arcsec. 

\begin{table*}\scriptsize
\centering
\begin{minipage}{140mm}
\caption{VLA observational parameters}
\label{4:tt1}
\begin{tabular}{@{}llrccccccrc@{}}
\hline
Field \footnote{Fields VLA--A and VLA--B   (D--array) correspond to VLA Project ID: AC857}&
${\alpha}_{2000}$ & ${\delta}_{2000}$ & Array &Integration & Beam\footnote{Beam size for the natural weight cubes}  &Channel\footnote{post Hanning smoothing the velocity resolution is 2 $\times$ channel separation}  &Velocity& rms & rms  \\
& & & config & time &size& separation &range & & \\
& [$^h$ $^m$ $^s$] & [\degree\ \prim\ \prin\ ] & & [hours]& [\prin]&
[\km ] & [\km ] & [mJy\,beam$^{-1}$ ]&[K]\\ \hline 
A& 11 37 01.0 & 20 02 08.8& D &2.7& 69 x 60 & 21 &5763--6859& 0.33 &0.05 \\ 
B &11 39 08.3 & 19 39 02.2 & D & 2.3 & 72 x 60 &21 &6261--7362& 0.36& 0.05 \\
 \hline
\end{tabular}
\end{minipage}
\end{table*}
\normalsize

\renewcommand{\thefootnote}{\alph{footnote}}
\begin{table*}
\begin{minipage}[h]{\linewidth}\centering
\begin{center}
\caption{ Properties of \hi\  detections}
\label{4:detections}
\begin{tabular}{@{}llllllllc@{}}
\hline 
  \multicolumn{1}{l}{ Field} & 
  \multicolumn{1}{l}{Galaxy ID} &
  \multicolumn{1}{l}{RA (2000)$^a$} &
  \multicolumn{1}{l}{Dec (2000)$^a$} &
  \multicolumn{1}{l}{Type$^b $} &
  \multicolumn{1}{l}{ V$_\mathrm{opt}^c$} &
   \multicolumn{1}{l}{V$_\mathrm{HI}$} &
  \multicolumn{1}{l}{W$_{20}$} &
   \multicolumn{1}{l}{M$_\mathrm{HI} $}\\
      \multicolumn{1}{l}{} & 
  \multicolumn{1}{l}{Identifier} &
  \multicolumn{1}{l}{[h m s]	} &
  \multicolumn{1}{l}{[\degree\ \prim \prin ]		} &
  \multicolumn{1}{r}{} &
  \multicolumn{1}{l}{[\km]} &
 \multicolumn{1}{l}{[\km]} &
 \multicolumn{1}{l}{[\km]} &
   \multicolumn{1}{c}{[10$^9$ \msolar]} \\
     \hline 
VLA--A	&MAPS--NGP O 434 2505 	&	11 36 49.61	&	20 09 40.5	&	S 		&	6296&	6285$\pm$5	&150$\pm$10	&	1.2	\\
VLA--A	&	MRK 0182			&	11 36 54.00	&	19 55 34.81	&	Compact	&	6328	&	6245$\pm$6	&\,\,\,70$\pm$2&	0.4	\\
VLA--A	&	\cg027				&	11 36 54.23	&	19 59 50.04	&	Sc  		&	6630&	6630$\pm$48	&240$\pm$96	&	1.2	\\
VLA--A	&	\cg026 (disk +tails)		&	11 36 54.40	&	19 58 15.00	&	SBa pec	&	6191	&	6195$\pm$2	&390$\pm$4	&	17.8	\\
VLA--A	&	SDSS J113655.45+200645.1	&	11 36 55.45	&	20 06 45.1	&	--	&	--	&	6120$\pm$9	&\,\,\,60$\pm$18&	0.2	\\
VLA--A	&	SDSS J113709.84+200131.0	&	11 37 09.84	&	20 01 31.0	&	--	&	6056	&	6140$\pm$11	&240$\pm$22	&	2.3	\\
VLA--A	&	AGC 210538			&	11 38 03.82	&	19 51 41.9	&	--		&	6196&	6195$\pm$13	&210$\pm$26	&	1.9	\\
VLA--B 	&	\cg036				&	11 38 50.98	&	19 36 05.24	&	S0/a		&	6787	&	6810$\pm$6	&100$\pm$12	&	0.8	\\
VLA--B 	&	FGC\,1287 (disk +tail)	&	11 39 10.90 	&	19 35 06.0	&	Sdm		&	6825	&	6780$\pm$2	&220$\pm$4	&	9.4	\\
VLA--B 	&	\cg041 				&	11 39 24.40	&	19 32 7.2 		&	Sb 		&	6778&	6778$\pm$5	&259$\pm$10	&	2.8	\\
Arecibo 	&	LBSC D571-03 		&	11 38 28.15	&	19 58 50.0	&	Sm 		&	6989&	6983$\pm$1	&124$\pm$2	&	1.8	\\
\hline
\footnotetext[1]{Optical position from NED.}
\footnotetext[2]{Hubble type from GOLDMine (Gavazzi et al. 2003b) \nocite{gava03b} or if unavailable NED.}
\footnotetext[3]{Optical velocity from NED or if unavailable from SDSS redshift.}

\end{tabular}
\end{center}
\end{minipage}
\end{table*}

\normalsize

Field VLA--A encompasses  the galaxy group RSCG 42 which has a velocity of 6296 \km\ and lies $\sim $ 108 arcmin (2.7 Mpc) West of the A\,1367 cluster centre. In this data set the \hi~mass detection threshold is \aprox 1$\times$10$^{8}$ M$_{\odot}$ (corresponding to 3$\sigma$ over 2 consecutive 21 \km\ channels). The equivalent column density sensitivity for emission filling the beam is then $\sim$ 2 $\times$10$^{19}$\,cm$^{-2}$. Field VLA--B contains the FGC\,1287 triplet.  The \hi\ mass detection threshold is the same as for field A.

We used the ALFA feed array at the 305--m Arecibo Telescope to observe \hi\ towards two positions near LSBC~D571-03 in a 5 and 10  minute on--source integration on June 8, 2011. LSBC~D571-03 is a candidate to have interacted with FGC\,1287; the original \hi\ detection of LSBC~D571-03 in AGES was heavily contaminated by RFI. 

\section{RESULTS}
\label{results}
The properties for each of the \hi\ detections in both observed VLA fields and in one of the Arecibo follow--up pointings are set out in Table \ref{4:detections}. The table shows the field, galaxy name, morphological type, optical velocity, \hi\ mass, \hi\ velocity, \hi\ velocity width, and \hi\ mass.

\medskip
\noindent
{\em Field VLA--A: RSCG\, 42} \
\medskip

The positions of the \hi\ detections in field VLA--A are
shown in the bottom right panel of  Figure  \ref{fig1}.  All detections,
except AGC\,210538, fall within the 32 arcmin FWHM of
the primary beam. The Redshift Survey of Compact Groups
 \citep[RSCG;][]{barton96} catalogues RSCG\,42 as a compact
group with 3 members (\cg026, \cg027 amd MRK\,0182). The most striking \hi\ feature is the $\sim $160 kpc tail extending
NE from \cg026 towards SDSS J113709.84+200131.0, with an indication of a smaller tail extending W of \cg026
which are consistent with a tidal interaction between these
two group members. The orientation of the tails is approximately
perpendicular to the major axis of \cg026.
Both galaxies are significantly bluer (SDSS \textit{g-i } = 0.2
and 0.21 respectively) than the other galaxies detected in \hi\ in this VLA field.

Based on the method from \cite{kew02} the
SFR(FIR) for \cg026 is $\sim$5.7 \msolaryr. The optical,
NIR and \halpha\ disks of both \cg026 and \cg027 ($\Delta V\sim400$\,\km)  are highly perturbed and the \hi\
velocity field (not shown) reveals \hi\ at intermediate velocities
between the two galaxies, providing confirmation that
\cg027 is involved in the interaction as well. The
complementary \hi\ deficiencies (-0.84 and 0.48 respectively)
suggests that \hi\ from \cg027 (an Sc spiral) has been
displaced in the interaction. We estimate the \hi\ mass of the tail to be $9.3 \pm 1 \times  10^9$\,\msolar.  Overall the distribution and
kinematics of the \hi\ is consistent with a tidal interaction
amongst the group members.\

\medskip
\noindent
{\em Field VLA--B:  FGC\,1287} \
\medskip

\hi\ was detected in FGC\,1287 (Sdm,  \textit{g-i} = 1.21) and two neighbouring galaxies (\cg041 and \cg036) with a remarkably small spread in \hi\ velocities (32 \km). Their
positions are shown in the lower left panel of Figure \ref{fig1}.
All three detections in \hi\, including the FGC\,1287 tail, lie
within the FWHM of the field's primary beam. \hi\ velocities
along the optical major axis of FGC\,1287 are consistent
with a rotating \hi\ disk with velocities in the range of
$\sim$ 6600 \km\ to 6800 \km. An additional velocity component along the major axis ($\sim$6900 \km) is located near
the NE edge of the optical disk and contains $\sim$ 25 \% of the \hi\ mass in the disk. \textcolor{black}{We estimate the \hi\ mass of the tail to be $5.7 \pm 1 \times  10^9$\,\msolar. } 

A faint \hi\ bridge, seen just below the 3$\sigma$ level,
seems to be joining the FGC1287 \hi\ tail to \cg036, suggesting an interaction.
However, no sign of a perturbed optical/\halpha\ morphology is observed in
either \cg036 or FGC\,1287. For FGC\,1287
$\log({\mathrm{F(H\alpha)}}) = -13.42 \pm 0.06$\,erg\,cm$^{-2}$\,s$^{-1}$, EW(\halpha) = $25.1\pm 3.5$\,\AA\ (Gavazzi, priv.\ comm.).  It is important
to remember that FGC\,1287 is almost perfectly edge-on, making
it difficult to clearly determine its optical morphology.


\begin{figure}
\begin{center}
\includegraphics[ angle=0,scale=.4] {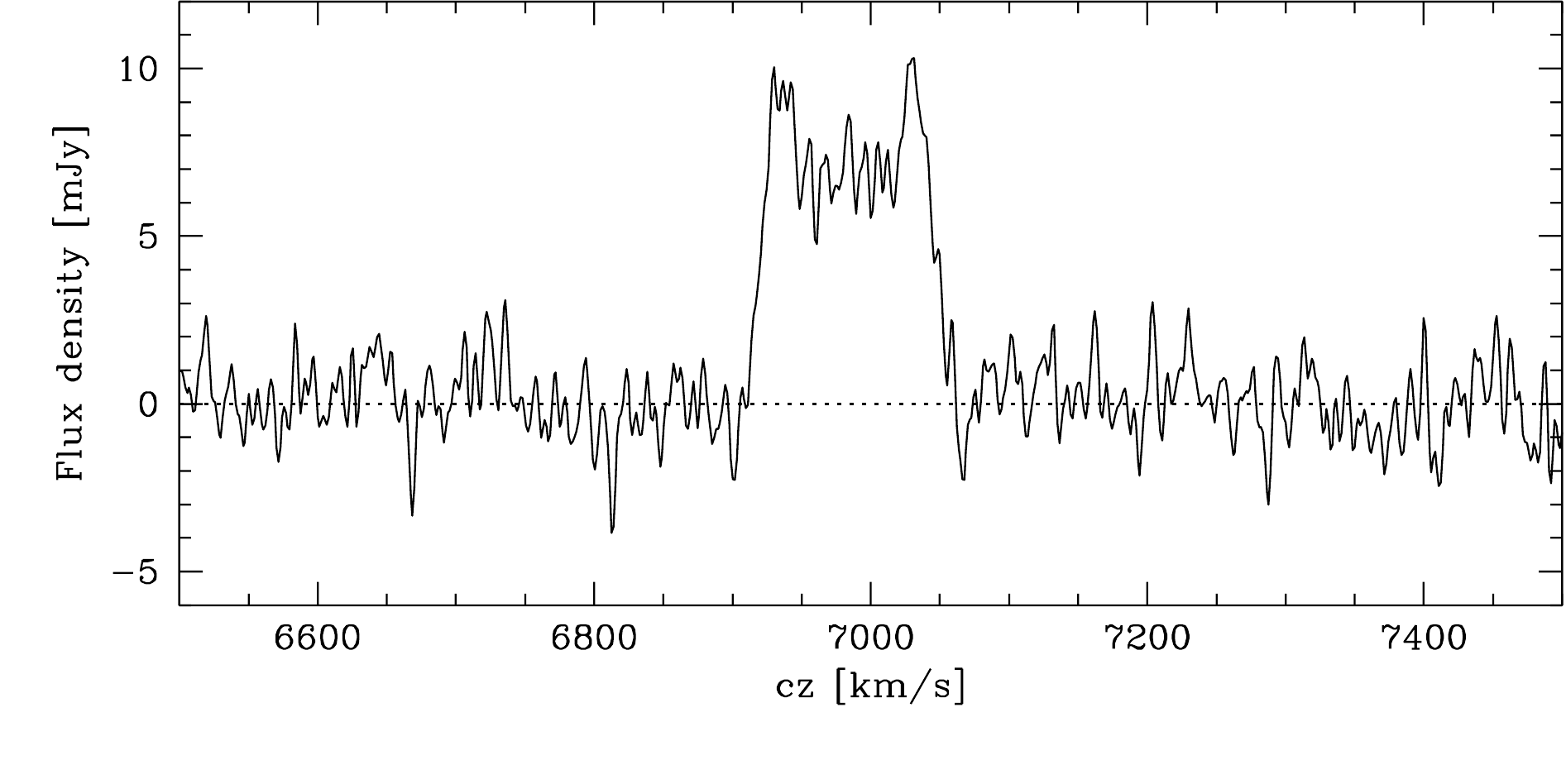}
\vspace{1cm}
\caption []{LSBC~D571--03 Arecibo \hi\ spectrum at the position of the optical galaxy. The velocity resolution is 4 \km; the   rms noise is 1.22 mJy. }
\label{ara}
\end{center}
\end{figure} 

\renewcommand{\thefootnote}{\arabic{footnote}}

\section{Discussion and concluding remarks}
\label{4:conclusion}
As the lower panels of Figure  \ref{fig1} clearly show, RSCG\,42 and FGC\,1287 both
contain spectacular \hi\ tails with projected lengths of $\sim$ 160
kpc and 250 kpc respectively. \textcolor{black}{The large \hi\ masses in these groups imply they are on their initial infall to the cluster. Had the groups previously transited the cluster core almost all the \hi\ would be expected to have been removed \citep{abadi99, bravo00}. Moreover, the cluster crossing time of $\sim2.5 \times 10^9$\,yr being much larger than the \hi\ tail survival time scale, together with the tail orientations, rules out a cluster core transit as the origin of the tails.} In both cases the \hi\ mass in
the tails is approximately equal to the \hi\ in the two galaxies
from which the tails appear to emanate (\cg026 and
FGC\,1287). Overall the available evidence seems to favour a
tidal interaction amongst the RSCG\,42 group members as
the cause of the \cg026 \hi\ tail.

For FGC\,1287 the cause of the tail is less clear.
Whereas the morphology of the tail could at a first sight resemble
similar features observed in ram-pressure stripped
cluster galaxies, the ICM densities extrapolated from the cluster X--ray
emission at the distance of this group would require unrealistically
high relative velocities and is inconsistent in direction with
infall to the cluster. Although we cannot exclude that the group harbours a hot
intra-group medium component, a hydrodynamical mechanism
would not be able to explain why only FGC\,1287 was subject
gas stripping, and cannot easily reproduce the change in direction
along the tail.

Tidal interactions amongst the low velocity dispersion triplet could
account for the large \hi\ mass which appears to have been
removed from the disk of FGC\,1287, but is inconsistent with the lack
of disturbed features in the other group members. More importantly,
the extraordinary length of the \hi\ tail suggests that the interaction forces operated
principally in the plane of the sky.
Moreover, the interaction which produced the \hi\ tail appears to have caused
remarkably little damage to the stellar and \halpha\ disk of FGC\,1287. This lack of stellar disk perturbation is more characteristic
of a high--speed velocity tidal interaction  than a  low--velocity gravitational
encounter amongst the group members.
Observations of other groups, such as Stephan's Quintet
show that high density groups with a high speed intruder
can have a large fraction of their \hi\ outside the disks of the
group members and lead to long tails with changes in direction
along the tails \citep{williams02,renaud10}, although in that case with damage to the optical disks and
associated stellar tails.
The only other \hi\ tail over 200 kpc long known to date
is the one associated with NGC\,4254 and,  in that case,
a high-speed interaction has emerged as the most likely scenario.
In order to look for possible candidate intruders we
used NED to search all galaxies within a $\pm$ 500 \km\ velocity range and 750\,kpc radius to find any additional bound group members and non--bound, high--velocity interaction partners.

Our search revealed a single low surface brightness galaxy,
LSBC~D571-03 (11h38m28.1s, +19d58m50s), V$_\mathrm{opt}$ = 6989 \km\ projected 25.1 arcmin (640 kpc) from
FGC\,1287. Intriguingly, this galaxy lies exactly on the continuation of
the \hi\ tail of FGC\,1287. The projected separation from FGC\,1287 is
consistent with an interaction with a motion in the plane of the sky of 1000
\km\ that happened $\sim$ 6 $\times$ 10$^8$ years ago; such a high speed encounter would
be compatible with the lack of optical disturbance in FGC\,1287.
In addition, the velocities at the end of the \hi\ tail (6820 \km) are
intermediate between those in the rotating disk of FGC\,1287
(6600 \km\ to 6750 \km) and the \hi\ detection of
LSBC~D571-03 (Figure \ref{ara}). Finally, the velocity of the anomalous \hi\ component in the
disk of FGC1287  (6850\,\km\ to 7000 \km) is similar to that of
LSBC~D571-03 (V$_\mathrm{HI}$ = 6983\,\km).
However, the optical morphology and \textcolor{black}{the Arecibo} \hi\ single dish spectrum (Fig.~\ref{ara})
do not show any obvious sign of disturbance.
Moreover, the fact that LSBC~D571-03 at m$_g$ = 16.4 appears to be significantly fainter than FGC\,1287 (m$_B = 14.31$)raises the question why only the brightest system is showing clear signs
of interaction. Thus, for the moment, we can only speculate that this tail
may have been produced by a high speed encounter with another galaxy
and only additional data will allow us to gain additional insights
into this unique system.

Whatever the true mechanism(s) at work here, our new discoveries highlight how
little we know about environmental effects in groups \textcolor{black}{infalling to clusters} and emphasize the importance of group
interactions in the evolution of cluster spirals beyond the central cluster region.

\section*{Acknowledgments}

We wish  to thank Giuseppe Gavazzi for kindly providing  \halpha\  data for FGC 1287.
\textcolor{black}{This work has been partially supported by research projects
AYA2008--06181--C02  from the Spanish Ministerio
de Ciencia y Educaci\'on and the Junta de Andaluc\'{\i}a (Spain) grants P08--FQM--4205 and TIC--114.} 
\textcolor{black}{HBA thanks CONACyT for financial support.} 

This research has made use of the NASA/IPAC Extragalactic Database (NED) which is operated by the Jet Propulsion Laboratory, California Institute of Technology, under contract with the National Aeronautics and Space Administration.

This research has made use of the Sloan Digital Sky Survey (SDSS). Funding for the SDSS and SDSS-II has been provided by the Alfred P. Sloan Foundation, the Participating Institutions, the National Science Foundation, the U.S. Department of Energy, the National Aeronautics and Space Administration, the Japanese Monbukagakusho, the Max Planck Society, and the Higher Education Funding Council for England. The SDSS Web Site is http://www.sdss.org/.

\bibliographystyle{mn2e}
\bibliography{cluster}

\end{document}